# Cloud based VANET Simulator (CVANETSIM)

Mohammad Mukhtaruzzaman and Mohammed Atiquzzaman
*School of Computer Science*
*University of Oklahoma*
Norman, OK-73019
mukhtar@ou.edu, atiq@ou.edu

*Abstract—* Vehicular ad hoc network (VANET) is an integral part of vehicular communication. VANET suffers many problems such as scalability. To solve scalability and other problems of VANET, clustering is proposed. VANET clustering is different than any other kind of clustering due to the high mobility of the vehicles. Likewise, VANET and VANET clustering, VANET simulator requires some unique features such as internet based real-time data processing, huge data analysis, the complex calculation to maintain hierarchy among the vehicles, etc.; however, neither web based VANET simulator nor clustering module available in the existing simulators. Therefore, a simulator that will be able to simulate any feature of VANET equipped with a clustering module and accessible via the internet is a growing need in vehicular communication research. At the Telecom and Network Research Lab (TNRL), University of Oklahoma, we have developed a fully functional discrete-event VANET simulator that includes all the features of VANET clustering. Moreover, the cloud based VANET simulator (CVANETSIM) is coming with an easy and interactive web interface. To our best of our knowledge, CVANETSIM is the first of its kind which integrates features of the VANET simulator, built-in VANET clustering module, and accessible through the internet.

I. INTRODUCTION

Internet of vehicles (IoVs) and connected vehicles require vehicular communication which is facilitated by the vehicular ad hoc network (VANET). VANET has some unique features such as high mobility that requires specialized software to simulate. Moreover, VANET also has some specialized protocols such as clustering [1, 2] which requires complex calculation for a hierarchical structure. To simulate a VANET scenario, specifically two parts are simulated. The first part is to generate traffic that is provided by traffic simulator like simulation for urban mobility (SUMO) [3], MOVE[4], etc. The second part is the main part of the VANET simulation what is simulated by a traditional network simulator or VANET simulator.

Internet-of-things (IoTs), internet-of-vehicles (IoVs), etc. are advancing rapidly. At the same time, simulation platforms for VANET are not advancing at a steady rate that can meet the growing demand for advanced simulation tools such as VANET clustering which requires the use of a large database, real-time data processing, complex multi-level hierarchical calculation, etc. Moreover, the platform should be accessible through the internet, preferably, a cloud-based platform with a user-friendly environment.

Some simulation platforms have been developed over the years; however, while VANET research is advancing rapidly, the simulation platforms are not advancing concurrently. For example, no simulator is getting more popularity than ns-2 while NS-2 has not been updated in a decade, since 2011[5]. Some attempts have been made to develop specialized VANET simulators such as TraNS [6], veins [7], Netsim [8], etc.; however, these platforms are either proprietary or need a specific operating system and specialized setup. Also, they do not have a clustering module or accessibility over the internet.

As part of our research, we needed a VANET simulation platform that is equipped with a fully functional clustering module, easy access of data through the database server, machine-independent, and easily accessible over the internet. Moreover, a cloud based VANET simulation platform will facilitate a path providing an application for real-life vehicular communication.

Our *objective* was to develop a fully functional VANET simulator with a complete clustering module. Our *aim* is to develop the simulator which will be using a database server for large data and will be accessible over the internet. Cloud based VANET simulator (CVANETSIM) will be the first of its kind. The following features will be available for CVANETSIM:
- All necessary features of a VANET simulator,
- Fully functional clustering module,
- Using a database to provide efficient access and update on data,
- Cloud based platform to access over the internet.

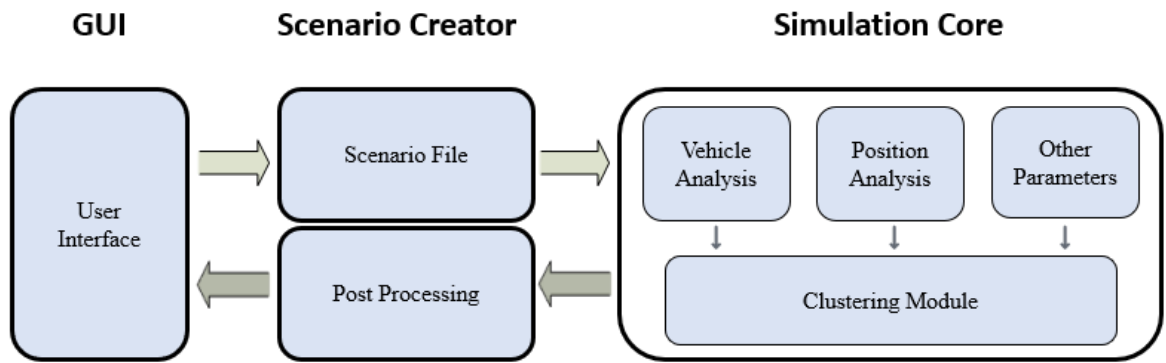

Fig. 1. Architecture of the VANET simulator, CVANETSIM.

In the next section (Section II), CVANETSIM has been described and Section III represents a conclusion with a future plan on CVANETSIM.

## II. CLOUD BASED VANET SIMULATOR (CVANETSIM)

The discrete-event cloud based VANET simulator, CVANETSIM, is developed using Java programming language and MySQL database in a Tomcat web server. A traditional network simulator was not developed keeping in mind the VANET scenario. VANET requires analysis of big data as well as clustering protocol demands complex multi-level calculation. Due to this combination, clustering protocols are not readily available in the traditional network simulator. Moreover, to build a real-life application for VANET, we need a web based VANET simulator that is capable of handling big data using a database. The simulator will have an efficient clustering protocol and can be accessed over the internet. CVANETSIM is developed especially for VANET scenarios and fulfilled all these requirements including a VANET clustering protocol. Since CVANETSIM is accessible from the internet, CVANETSIM does not serve as a simulator only, any kind of real-life VANET application can be developed using this simulator without storing or processing any data on a local machine. CVANETSIM is a discrete-event simulator which uses MySQL database to use SUMO data and then process data to generate a cluster. CVANETSIM processes SUMO data and analyzes various features of vehicles such as degree, transmission range, velocity, relative velocity, distance, position, relative distance, angle, etc. After extracting all information, the cluster is formed, and the CH is selected based on an individual algorithm.

The architecture of the VANET simulator, CVANETSIM, is shown in Fig. 1. CVANET has four main components: GUI, Scenario file, Simulation core, and post-processing engine. GUI is the graphical user interface of CVANETSIM. The scenario file is the traffic data given as the input. An analysis is performed on features of vehicles, the position of the vehicles, and other necessary parameters such as features of dedicated short-range communication (DSRC), etc. If clustering protocol is needed to be implemented, the clustering module receives all analyzed data and process further to create cluster and cluster head (CH) depending on a particular algorithm and transmission range. Next, in the next processing phase, all other vehicles are assigned a cluster depending on the clustering protocol and transmission

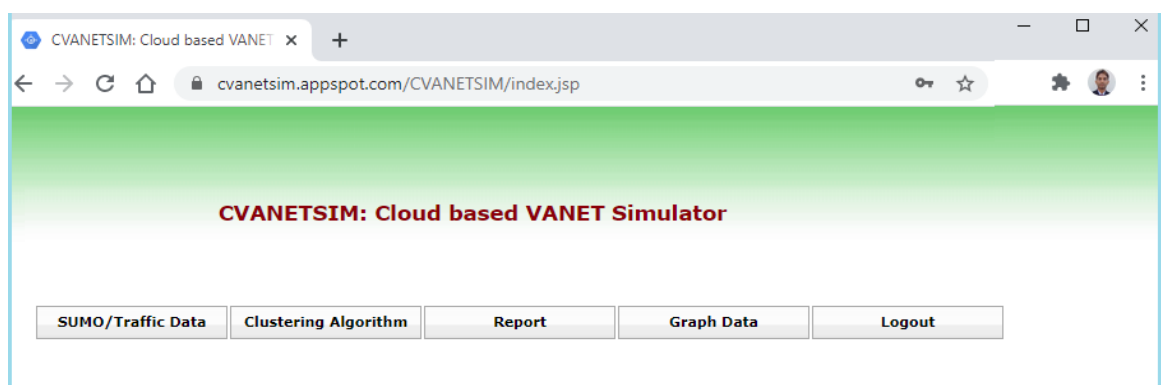

Fig. 2. Main menu items of the cloud based VANET simulator, CVANETSIM.

range. A prototype of the developed CVANETSIM is available on Google cloud [9].

GUI is the graphical user interface of CVANETSIM where the user can find some menu items after login. A screenshot of the developed CVANETSIM is shown in Fig. 2. Four main items in the menu are:

i. SUMO/Traffic Data: Data from SUMO or any traffic simulator are given as a scenario file as the input to be processed for the next phase. Vehicle features, location, other parameters are analyzed for every discrete event.

ii. Clustering Algorithm: Here is the code for a specific VANET clustering algorithm. Based on the algorithm cluster is formed and CH is assigned at this stage.

iii. Report: A detailed report is generated at this stage for each and every discrete event.

iv. Graph Data: From the processed data, necessary information is generated here that can be used to generate graph. CVANETSIM can generate the average CH duration, the average cluster member (CM) duration, the average CH change, the average number of clusters, the average number of CM, the average number of non-clustered vehicles etc.

III. CONCLUSSION

In this paper, we presented a novel and unique VANET simulator, CVANETSIM. The simulator can simulate any VANET scenario. Built-in clustering module can create clusters from any number of vehicles for any scenario depending on the clustering algorithm. This cloud-based simulator is accessible over the internet and comes with interactive and user-friendly GUI.

In future works, we want to integrate more and more VANET protocols and to implement an android version for mobile application.

**Mohammad Mukhtaruzzaman** received the BS and MS in Computer Science and Engineering from Khulna University of Engineering (KUET) and Technology, and Bangladesh University of Engineering and Technology (BUET). He is pursuing Ph.D. in Computer Science at University of Oklahoma (OU), USA. His research interests include VANET, ITS, wireless and mobile networks, machine-learning for IoT and big data for financial systems. More information: www.mukhtaruzzaman.com

**Mohammed Atiquzzaman** holds the Edith Kinney Gaylord Presidential professorship at the University of Oklahoma. He is the editor-in-chief of Journal of Network and Computer Applications, founding editor-in-chief of Vehicular Communications, associate editor of many journals including IEEE journals. His research has been funded by NSF, NASA, US Air Force, Cisco, Honeywell, etc. His publications can be found at www.cs.ou.edu/~atiq